\documentclass[usenatbib,usegraphicx,twocolumn,a4paper]{mn2e}
\usepackage[latin1]{inputenc}
\usepackage{graphicx,amsmath}
\usepackage{times}
\usepackage{amssymb}
\usepackage{natbib}
\usepackage{rotating}
\usepackage{lscape}
\usepackage{multirow}
\usepackage{ulem}

\graphicspath{{images/}}

\title[Imprints of Dark Energy on Halo Profiles]{Imprints of Dark Energy on Cosmic Structure Formation: III) Sparsity of Dark Matter Halo Profiles}
\author[I.~Balm\`es et al.]{I.~Balm\`es\thanks{E-mail: irene.balmes@obspm.fr}, Y.~Rasera, P.-S.~Corasaniti, and J.-M.~Alimi\\
CNRS, Laboratoire Univers et Th\'eories (LUTh), UMR 8102 CNRS, Observatoire de Paris,
Universit\'e Paris Diderot,\\ 5 Place Jules Janssen, 92190 Meudon, France}
\begin{document}

\date{}

\maketitle

\label{firstpage}

\begin{abstract}
We study the imprint of Dark Energy on the density profile of Dark Matter halos using a set of high-resolution large volume cosmological N-body simulations from the Dark Energy Universe Simulation Series (DEUSS). We first focus on the analysis of the goodness-of-fit of the Navarro-Frenk-White (NFW) profile which we find to vary with halo mass and redshift. We also find that the fraction of halos ill-fitted by NFW varies with cosmology, thus indicating that the mass assembly of halos with perturbed density profiles carries a characteristic signature of Dark Energy. To access this information independently of any parametric profile, we introduce a new observable quantity: the halo sparsity $s_\Delta$. This is defined as the mass ratio $M_{200}/M_\Delta$, i.e. the ratio of mass inside a sphere of radius $r_{200}$ to that contained within a radius $r_\Delta$, enclosing $200$ and $\Delta$ times the mean matter density respectively. We find the average sparsity to be nearly independent of the total halo mass, while its value can be inferred to better than a few percent from the ratio of the integrated halo mass functions at overdensities $\Delta$ and $200$ respectively. This provides a consistency relation that can validate observational measurements of the halo sparsity. Most importantly, the sparsity significantly varies with the underlying Dark Energy model, thus providing an alternative cosmological probe.
\end{abstract}

\begin{keywords}

\end{keywords}

\section{Introduction}
In the standard cosmological scenario initial
Dark Matter (DM) density fluctuations are the seeds of the cosmic structures we observe today.
Gravitational instability amplifies these perturbations that trigger
the collapse of the baryonic gas. At early times and on the large scales the dynamics of this
process is linear. In contrast at late times and on small scales, as fluctuations grow sufficiently large,
the gravitational collapse becomes highly non-linear. It is during this phase that DM particles eventually virialize into 
gravitationally bounded objects, the halos. 

In the hierarchical bottom-up scenario
low mass halos form earlier, while massive ones are assembled
at later times through mergers of smaller mass halos and accretion of DM particles from 
the surrounding density field. A complete understanding of this regime is key to disclose the 
processes that shapes the distribution of matter in the universe. It is inside halos that 
cooling baryonic gas falls in to form the stars and galaxies that surround us. Moreover, these carry 
cosmological information that can be tested through galaxy survey observations. 

In the future a new generation 
of survey experiments will probe the nature of the invisible components in the universe through 
accurate measurements of the clustering of matter on an unprecedented range of scales. This calls 
for a major theoretical effort to provide reliable cosmological model predictions. However, 
because of the complexity of the gravitational collapse, cosmological studies of the DM clustering 
have mainly relied on numerical N-body simulations. 

Numerous works have been dedicated to studying the imprint of Dark Energy on the non-linear 
cosmic structure formation. In this series of papers we have studied the signature that DE leave on the non-linear 
matter power spectrum and the halo mass function. In \citet{Alimi2010} we have shown that DE alters the non-linear 
clustering of Dark Matter at small scales in a very peculiar manner. 
This is a direct consequence of the fact that DE affects the linear growth of matter density fluctuations and 
on scales above the stable clustering regime the non-linear collapse carries an integrated record of the past 
linear evolution \citep[see e.g.][]{Ma2007}. Similarly, the study of the halo mass function in \citet{Courtin2011} 
has shown that deviations from a universal multiplicity function strongly correlate with 
the critical density threshold and the virial density predicted by the spherical collapse model of the simulated cosmology. 
In this paper we specifically focus on the imprint of DE on the density profile of DM halos.

One remarkable result of N-body simulations studies is that Dark Matter halos, regardless of the mass or the 
characteristics of the underlying cosmological model, exhibit a universal density profile which can be described in 
terms of a two-parameter fitting formula, the so called Navarro-Frenk-White (NFW) profile \citep[][]{Navarro1995,Navarro1996}. 
The cosmology dependence is entirely encoded in the relation between the NFW parameters, that is the mass dependence 
of the concentration parameter. This provides a measure of the compactness of the halo as function of its mass, and 
has important observational implications, since measurements of the concentration of 
galaxy clusters can test cosmology and constrain the cosmological parameters. 

The mass and redshift dependence of the halo concentration has been studied in a vast literature 
\citep[see e.g.][]{Navarro1997,Bullock2001,Eke2001,Shaw2006,Neto2007,Duffy2008,Gao2008,Prada2012}. These studies have focused on the $\Lambda$CDM cosmology, while \citet{Dolag2004} first investigated the evolution of the concentration in non-standard Dark Energy models. Their analysis has shown that variations of the concentration-mass relation as function of redshift are related to differences of the linear growth rate of the underlying Dark Energy cosmologies. More recently, \citet{DeBoni2013} have found the slope of the $c-M$ relation to be roughly identical in all models, while the normalization depends on the linear growth rate.
 
Although the emergence of a universal halo density profile is still not understood, a number of empirical studies have suggested that 
the dependence of the concentration on halo mass and indirectly the appearence of the NFW profile is well correlated with the mass 
accretion history of halos \citep[see e.g.][]{Wechsler2002,Zhao2003,Zhao2009,Ludlow2013}.

On the observational side measurements of the concentration in massive clusters are still far from providing
conclusive results \citep{Buote2007,Schmidt2007,Comerford2007,Okabe2010,Ettori2010,Wojtak2010,Oguri2012}. 
One complication arises from the fact that massive halos may not be relaxed and consequently their profiles may not be smooth. In fact, it is not at all implausible that a large fraction of the most massive clusters consists of 
unrelaxed halos \citep{Ludlow2012}. Though the profile may depart from
the NFW formula (or any other parametric form of the profile) the mass distribution inside such halos still carry cosmological information. 

How can we access such information independently of the profile? Does Dark Energy leaves a distinctive signature on halos with perturbed density profiles? 

It is the goal of the work presented here to answer these questions. Using a set of numerical N-body simulations of different DE cosmologies
we perform a detailed study of the density profiles of DM halos. As a case study, we show that the fraction of halos which are poorly fit by the NFW 
varies substantially with cosmology. To make use of this effect we introduce the sparsity, an observable measure of the mass distribution in halos
which is independent of the halo density profile. We find this to be weakly dependent on the total halo mass, while it carries a distinct imprint of Dark Energy.
We show that on a sample of halos the average value of the sparsity is directly
related to the halo mass function, thus providing a self-consistent cosmological test 
applicable to all halos independently of the shape of their profile.

The paper is organized as follows: in Section \ref{nsim} we briefly describe the cosmological N-body simulations 
and the algorithms used to perform the numerical analysis; in Section \ref{fitnfw} we discuss the halo profile fitting
procedure and in Section \ref{chi2} we present the results of the NFW analysis. In Section \ref{spars} we introduce the sparsity and discuss
its relevant properties. In Section \ref{cosmospar} we describe its use as probe of cosmology and present our conclusions in Section \ref{conclu}.

\section{N-body Simulations}\label{nsim}
\subsection{Simulation sets}
We use a subset of N-body simulations from the ``Dark Energy Universe Simulation Series'' (DEUSS) and publicly available through the ``Dark Energy Universe Virtual Observatory''
(DEUVO) database\footnote{http://www.deus-consortium.org/deuvo/}. For more details on these simulations we refer the interested reader to dedicated sections
in \citet{Alimi2010,Rasera2010,Courtin2011}. These have been realized using the adaptive mesh refinement
code RAMSES based on a multigrid Poisson solver \citep{RAMSES,Guillet2011} for Gaussian initial conditions generated using the Zel'dovich approximation 
with MPGRAFIC code \citep{MPGRAFIC} and input linear power spectrum from CAMB \citep{CAMB}. All simulations have the same phase of the initial conditions.

We consider two class of cosmological models. \textit{Realistic} models, with parameters calibrated against measurements of the Cosmic Microwave Background anisotropies 
from the \textit{Wilkinson Microwave Anisotropy Probe} (WMAP) 5-year data \citep{Komatsu2009} and luminosity distances to Supernova Type Ia from the UNION 
dataset \citep{Kowalski2008}. These models include a standard flat $\Lambda$ Cold Dark Matter cosmology ($\Lambda$CDM-W5) and two quintessence scalar field models characterized by a 
Ratra-Peebles potential \citep[RPCDM-W5,][]{RatraPeebles1988} and supergravity inspired model \citep[SUCDM-W5,][]{Brax2000}. 
\textit{Toy} models are flat cosmological models with different background expansion and linear growth of
the density perturbations. We additionally require these models to have the same distribution of linear density fluctuations at
$z=0$, hence the same $\sigma_8$ value.
These include a large cosmological constant model (L-$\Lambda$CDM) with $\Omega_\Lambda=0.9$, a Ratra-Peebles quintessence model 
with large value of the slope of the scalar potential (L-RPCDM) and a Cold Dark Matter scenario (SCDM*, the * symbol is to remind that the model parameter values assumed here differ from the SCDM usually considered in the
literature; see e.g. \cite{Jenkins1998}). These
are used only for the purpose of studying the physical imprint of the underlying cosmological model on the Dark Matter halo profile. We also consider two $\Lambda$CDM models best 
fitting WMAP 1-year ($\Lambda$CDM-W1) and 3-year data ($\Lambda$CDM-W3) which have nearly identical linear growth histories and mainly differ for the value of $\sigma_8$. For all models
the reduced Hubble constant is set to $h=0.72$, apart $\Lambda$CDM-W3 for which $h=0.73$. The cosmological parameters of the simulated models are listed in Table~\ref{wmap_cosmo}. 

\begin{table} 
\begin{center}
\caption{\label{wmap_cosmo} Cosmological parameter values of the simulated models. $\Omega_{DE}$: density parameter for the Dark Energy component; $\sigma_8$: root-mean-square of fluctuations at the 8~Mpc/$h$ scale; $\alpha$: slope of the potential of the quintessence field; $\Omega_b$: density parameter for the baryons; $n_s$: scalar spectrum power-law index.} 
\begin{tabular}{cccccc}
\hline 
Model & $\Omega_{DE}$ & $\sigma_8$ & $\alpha$ & $\Omega_b$ & $n_s$ \\
\hline
$\Lambda$CDM-W5&0.74& 0.79& 0 & 0.044& 0.963\\
RPCDM-W5&0.77&0.66& 0.5 & '' & ''\\
SUCDM-W5&0.75&0.73& 1 & '' & ''\\
L-$\Lambda$CDM&0.9&0.79& 0 & '' & ''\\
L-RPCDM&0.74&0.79&10 & '' & ''\\
SCDM*&0&0.79&-& '' & ''\\
$\Lambda$CDM-W1&0.71& 0.90& 0& 0.047 & 0.99\\
$\Lambda$CDM-W3&0.76& 0.74& 0& 0.042 & 0.951\\
\hline
\end{tabular} 
\end{center}
\end{table}

We use data from simulations with $162\,h^{-1}\,\textrm{Mpc}$ boxlength and $512^3$ particles, and in the case of the realistic models we also
use data from simulations with $648\,h^{-1}\,\textrm{Mpc}$ boxlength and $1024^3$ particles. The characteristics of these simulations are summarized in Table \ref{simu_set}. 

\begin{table} 
\begin{center} 
\caption{\label{simu_set} Parameters of the N-body simulations
  for the various cosmological models: $z_i$ is
  the initial redshift, $N_{part}$ is the number of particles in the simulation,
$L$ is the simulation boxlength in units of h$^{-1}$Mpc, $m_p$ is the mass of the particle in units of h$^{-1}$M$_\odot$ and $\Delta$ the
  comoving resolution in units of h$^{-1}$kpc. All simulations share the same realization
  of the initial conditions, and
  start at high redshift (with $\sigma_{start} \simeq 0.05$ at the scale of the
  resolution of the coarse grid). Our refinement strategy consist in refining
  when the number of particles in one cell is greater than 8.} 
\begin{tabular}{cccccc}
\hline
Model&$z_i$&$N_{part}$&L&$m_p$&$\Delta$x\\
\hline
\textrm{$\Lambda$CDM-W5}&93&$512^3$&162&$2.29\times 10^9$&2.47\\
''&''&$1024^3$&648&$1.83\times 10^{10}$&9.89\\
\textrm{RPCDM-W5}&81&$512^3$&162&$2.02\times 10^9$&2.47\\
''&''&$1024^3$&648&$1.62\times 10^{10}$&9.89\\
\textrm{SUCDM-W5}&92&$512^3$&162&$2.20\times 10^9$&2.47\\
''&''&$1024^3$&648&$1.76\times 10^{10}$&9.89\\
\textrm{$\Lambda$CDM-W1}&93&$512^3$&162&$2.55\times 10^9$&2.47\\
\textrm{$\Lambda$CDM-W3}&''&''&''&$2.11 \times 10^9$&''\\
\textrm{L-$\Lambda$CDM}&''&''&''&$8.79 \times 10^8$&''\\
\textrm{L-RPCDM}&''&''&''&$2.29 \times 10^9$&''\\
\textrm{SCDM*}&''&''&''&$8.79 \times 10^9$&''\\

\hline
\end{tabular}
\end{center}
\end{table}

\subsection{Halo Finder Algorithms}
We detect halos using the Spherical Overdensity (SO) algorithm \citep{LaceyCole1994}. 
We estimate the density in each cell by counting the number of particles it contains, then the halo finder starts in the cell with the maximum density. 
The position of the center in the candidate cell is given by the particle with the largest number of neighbors in a sphere of given radius. The SO finder then draws spheres of increasing radii around that particle.
A halo is detected when the density $\rho$ enclosed in a given sphere is $\Delta$ times
the mean matter density $\rho_m$, with $\Delta$ the input parameter of the SO halo finder. 
In contrast, the Friend-of-Friend (FoF) algorithm \citep{Davis1985} detects halos as group of particles with an intraparticle 
distance smaller than an input linking-length parameter $b$. The advantage of this algorithm is that it does not impose any geometrical symmetry on 
the detected halos, although it tends to link bridged halos. 
In Appendix~\ref{fofhalos} we will show that results on the halo profile and halo sparsity obtained using FoF are in agreement with those
inferred from SO halos. On the other hand, we find that relevant differences manifests in the evaluation of the dynamical properties of the
detected halos.

Hereafter, we consider only halos with a minimum number of particles of $N_{min}=1000$. For a fair comparison of the halo properties among the different models we consider the same objects in the different simulations. This task is facilitated by the fact that we have set the same phase of the initial conditions, thus causing structures to form at the same positions in the simulation box.

\begin{figure*}
\begin{minipage}{17cm}
\begin{center}
\begin{tabular}{cc}
  \includegraphics{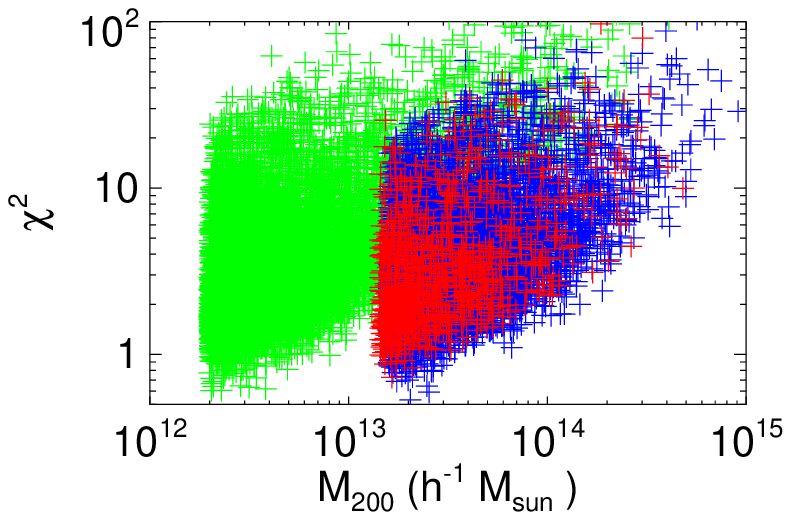}&
  \includegraphics{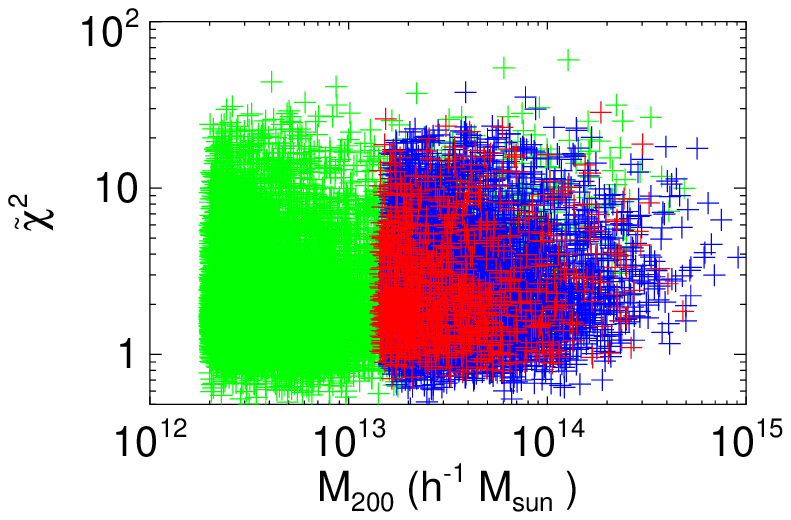}
\end{tabular}
  \caption{\label{convergence}Left panel: $\chi^2$-values for halos in the $256^3$ particle simulation catalog (red cross), $512^3$ (green cross) and $1024^3$ (blue cross) 
respectively as function of halo mass. Right panel: same plot for $\tilde{\chi}^2=\chi^2\sqrt{N_\mathrm{min}/N_\mathrm{part}}$-values. The dependence on the mass resolution 
has been largely reabsorbed in the rescaling of $\chi^2$.}
\begin{tabular}{cc}
  \includegraphics{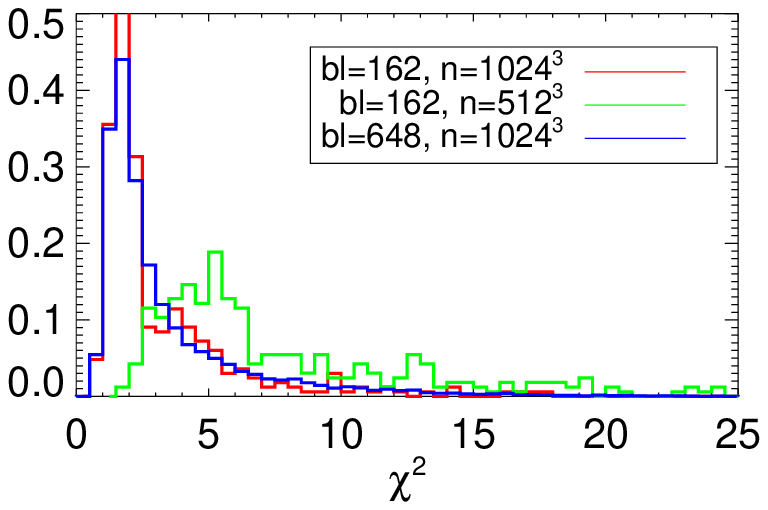}&
  \includegraphics{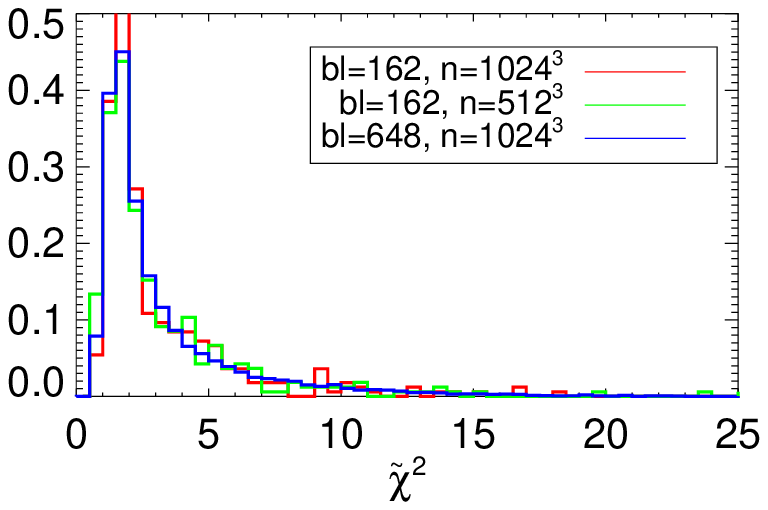}
\end{tabular}
  \caption{\label{convergence_hist}$\tilde{\chi}^2$-distribution for halos in the mass bin $1.6\times 10^{13}< \rmn{M}_{200} [\rmn{h}^{-1} \rmn{M}_{\sun}] < 2.0\times 10^{13}$ from 
162~h$^{-1}$ Mpc boxlength with $256^3$ (red histogram) and $512^3$ (green histogram) particles, and 648~h$^{-1}$ Mpc boxlength simulation with 
$1024^3$ (blue histogram) particles respectively.}
\end{center}
\end{minipage}
\end{figure*}

\section{Halo Density Profile}\label{fitnfw}
\subsection{Fitting Procedure}
The NFW profile \citep{Navarro1995} can be explicitly written in terms of the radius $r_{200}$ enclosing an overdensity $\Delta=200$ relative to the cosmic mean matter density, 
the enclosed mass $M_{200}$ and the concentration parameter $c=r_{200}/r_s$, where $r_s$ is the scale radius of the halo:
\begin{equation}
\rho_\rmn{NFW}(r) = \frac{M_{200}}{4\pi[\ln(1+c)-c/(1+c)]}\times \frac{1}{r \left( \frac{r_{200}}{c} +r \right)^2}.
\end{equation}

We fit the NFW formula to each halo in the simulation catalogs and define a $\chi^2$-statistics as
\begin{equation}\label{chi2stand}
\chi^2=\frac{1}{N_\rmn{bins}}\sum^{N_\rmn{bins}}_{i=1} \frac{[\rho_i-\rho_\rmn{NFW}(r_i)]^2}{2\sigma_i^2}, 
\end{equation}
where $N_\rmn{bins}$ is the total number of concentric shells binning the halo density profile, $\rho_i=n_i/V_i$ is the density in the $i$-th shell between radius $r_{i-1}$
and $r_i$, occupying a volume $V_i$ and containing $n_i$ particles; $\sigma_i=\sqrt{n_i}/V_i$ is the Poisson error. We discard all shells that do not contain
particles. Since the core of the halo is not well resolved we introduce a cut-off $\alpha$ and fit the NFW profile over the range $(\alpha\,r_{200},r_{200})$.
To perform the fit we fix the halo mass $M_{200}$ and $r_{200}$ to the values determined by the SO finder and then minimize Eq.~(\ref{chi2stand}) as function of the concentration parameter. 

Prior to discussing the physical implications of the NFW goodness-of-fit, we test for potential source of errors that may affect 
the evaluation of the $\chi^2$. In particular, we find the distribution of $\chi^2$ values systematically vary with the mass resolution of the simulations. To correct for this effect we introduce an alternative measure of the goodness-of-fit and verify that the inferred statistical distribution is independent of the fitting procedure. We limit this analysis to the $\Lambda$CDM-W5 model, but we have checked that the results apply to all simulated models.

\subsection{Numerical Tests}
\subsubsection{Mass Resolution}
In order to test for the effect of the mass resolution we consider an additional simulation with $162\,h^{-1}\,\textrm{Mpc}$ boxlength and $256^3$ particles which has the same mass resolution of the $648\,h^{-1}\,\textrm{Mpc}$ simulation with $1024^3$ particles. Halos of same mass in the $256^3$ simulation will be less resolved than those obtained in the $512^3$ run. Similarly for those in the $1024^3$ box compared to the $512^3$ case. Hence, in a given mass bin the $\chi^2$-statistics may vary with the resolution of the simulations, whereas it should be the same. In the left panel of Figure~\ref{convergence} we plot the $\chi^2$ of each SO halos as function of the halo mass $M_{200}$ for the $256^3$ (red cross), $512^3$ (green cross) and $1024^3$ (blue cross) simulations respectively. 
We can clearly see a systematic trend as function of halo mass and the resolution of the simulations. We find the effect to approximately scale as $\sqrt{N_\rmn{part}}$, where $N_\rmn{part}$ is the number of particles in each halo. To correct for this trend we rescale Eq.~(\ref{chi2stand}) as 
\begin{equation}\label{chi2tilde}
\tilde{\chi}^2=\chi^2\sqrt{\frac{N_\mathrm{min}}{N_\mathrm{part}}},
\end{equation}
which we plot in the right panel of Figure~\ref{convergence}. We can see that $\tilde{\chi}^2$ has absorbed most of the dependence on mass resolution. 
This can be better seen in Figure~\ref{convergence_hist} where we plot the normalized histogram of $\chi^2$ (left panel) and $\tilde{\chi}^2$ (right panel) for halos in a mass bin $1.6\times 10^{13}< \rmn{M}_{200} [\rmn{h}^{-1} \rmn{M}_{\sun}] < 2.0\times 10^{13}$ which is common to all three simulations. We can see that the probability distribution of $\chi^2$ is identical for the simulations with the $256^3$ and $1024^3$ particles, while it is different in the $512^3$ case. Using the rescaled variable $\tilde{\chi}^2$ the probability distributions are all statistically consistent, though we may notice small differences at the level of the peak and the tail of the $\tilde{\chi}^2$-distribution for the $512^3$ case. This may indicate the presence of some residual mass resolution effects, though 
differences are less than a few percent, hence smaller than the amplitude of the effects that will be discussed in the next sections and
thus negligible. Hereafter, we will use $\tilde{\chi}^2$ as a measure of the goodness-of-fit. However, before proceeding further let us make a few remarks that may help the reader to avoid any confusion related to our definition of $\tilde{\chi}^2$. We have introduced the scaling of $\chi^2$ by $\sqrt{N_{\mathrm{part}}}$ to account for the mass resolution effect, instead we have found convenient to normalize $\tilde{\chi}^2$ such that the lowest mass halos (for which $N_\mathrm{part}=N_\mathrm{min}$) have the same $\tilde{\chi}^2$ value independently of the simulation mass resolution (this is the $\sqrt{N_{\mathrm{min}}}$ factor in Eq.~(\ref{chi2tilde})). Though this may seem arbitrary it does not affect the analysis of the goodness-of-fit. In fact, a different choice of $N_\mathrm{min}$ changes the value of $\tilde{\chi}^2$ corresponding to a rigid shift along the y-axis of the points in Figure~\ref{convergence}. However, this overall normalization does not change the shape of the $\tilde{\chi}^2$-distribution. This is the key point since we do not use the absolute value of $\tilde{\chi}^2$ as indicative of the goodness-of-fit of NFW, but only as a relative measure. As we will discuss next, we are interested in defining populations of halos in terms of their probability of being well fit by NFW and this is quantified by the probability density distribution $\Pr(\tilde{\chi}^2)$ in terms of the relative difference $\Delta\tilde{\chi}^2$ and not the absolute value of $\tilde{\chi}^2$.

\subsubsection{Binning and Core Radius}
We now consider the effect of varying $N_\rmn{bins}$ and $\alpha$ on the goodness-of-fit, using the $162~h^{-1}$~Mpc boxlength and $512^3$ particles simulation. As a diagnostic we use 
the cumulative distribution function
\begin{equation}
\rmn{Q}(\tilde{\chi}^2_0) = \Pr(\tilde{\chi}^2 > \tilde{\chi}^2_0) = 1-\int_0^{\tilde{\chi}^2_0} \Pr(\tilde{\chi}^2) \rmn{d}\tilde{\chi}^2,
\end{equation}
which gives the fractional percent of halos whose $\tilde{\chi}^2 > \tilde{\chi}^2_0$. 

 We have computed $\rmn{Q}(\tilde{\chi}^2_0)$ for several values of $N_\rmn{bins}$ with $\alpha=0.01$ and found no significant variation provided the number of radial bins is sufficiently large, $N_\rmn{bins}>12$. On the other hand the fitting procedure is more sensitive to the choice of the core radius. We have determined the cumulative distribution for different values of $\alpha$ with $N_\rmn{bins}=60$. We find that removing a large fraction of the core radius ($\alpha>0.1$) greatly alters $\rmn{Q}(\tilde{\chi}^2_0)$, while too small values of $\alpha$ result in a fit that is sensitive to the poor resolution of the halo core, thus causing systematically larger values of $\tilde{\chi}^2_0$. We find a good compromise between these competing effects for $\alpha=0.1$. Thus, for the fitting procedure we consider $N_\rmn{bins}=60$ bins spaced logarithmically and $\alpha=0.1$. Hereafter, we will use these values
unless specified otherwise.

\section{NFW Profile and Goodness-of-Fit}\label{chi2}
We now focus on the goodness-of-fit of the NFW profile as function of the halo properties, using both the $162~h^{-1}$~Mpc boxlength, $512^3$ particles and the $648~h^{-1}$~Mpc boxlength, $1024^3$ particles simulations. As already mentioned the cumulative distribution function $\rmn{Q}(\tilde{\chi}^2_0)$ is
a useful diagnostic since it provides us with a quantitative estimate of the fraction of halos with $\tilde{\chi}^2 > \tilde{\chi}^2_0$. More specifically, from $\rmn{Q}(\tilde{\chi}^2_0)$ we can
classify the halo population according to the probability that their profile is fitted by the NFW profile. We find that halos with $\tilde{\chi}^2\lesssim 3$ are within $1\sigma$ ($68\%$ probability)
of the NFW profile, while those with $\tilde{\chi}^2\gtrsim 10$ are poorly fit at more than $2\sigma$ ($95.5\%$). In Table~\ref{table:cdf} we report the exact value of $\tilde{\chi}^2_0$
corresponding to the $1$ and $2\sigma$ limits for each simulated model at $z=0$ and $1$ respectively.

\begin{table}
\begin{center}
\caption{\label{table:cdf}Values of $\tilde{\chi}^2$ corresponding to 1 and 2 $\sigma$ deviation from the NFW profile for the simulated cosmologies at $z=0$ and $1$ respectively.}
\begin{tabular}{ccccc}
\hline
& \multicolumn{2}{c}{$z=0$} & \multicolumn{2}{c}{$z=1$} \\
& 1-$\sigma$ & 2-$\sigma$ & 1-$\sigma$ & 2-$\sigma$ \\ \hline
$\Lambda$CDM-W5 & $3.69$ & $12.00$ & $3.30$ & $8.78$ \\
RPCDM & $3.82$ & $11.47$ & $3.37$ & $9.08$ \\
SUCDM & $3.84$ & $11.69$ & $3.33$ & $9.17$ \\
L-$\Lambda$CDM & $4.61$ & $15.39$ & $3.32$ & $9.85$ \\
SCDM* & $3.05$ & $9.63$ & $3.24$ & $8.74$ \\
\hline
\end{tabular}
\end{center}
\end{table} 

\begin{figure*}
\begin{minipage}{18cm}
\begin{center}
  \includegraphics{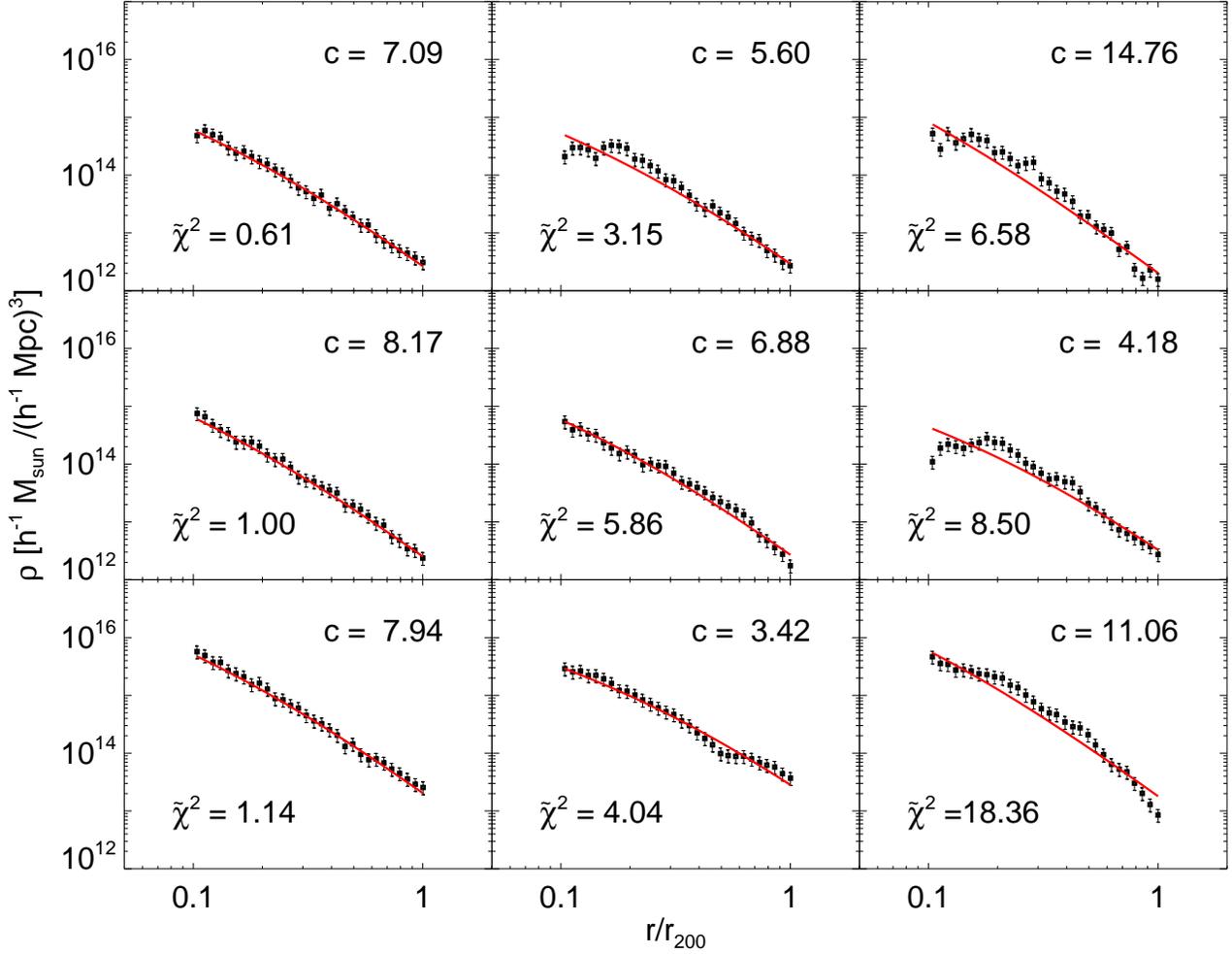}
  \caption{\label{fits}Density profiles of nine randomly selected individual $\Lambda$CDM-W5 halos in mass bins 
$10^{12}<M_{200}(\rmn{h^{-1}M_\odot})<10^{13}$ (top panels), 
$10^{13}<M_{200}(\rmn{h^{-1}M_\odot})<10^{14}$ (middle panels) and 
$M_{200}>10^{14}$ h$^{-1}\rmn{M_\odot}$ (bottom panels) with $\tilde{\chi}^2$ near the minimum value (left panels), 
at $\sim 1\sigma$ (central panels) and $\gtrsim 2\sigma$ (right panels) respectively. The red solid line is the best-fit NFW profile
in the range ($0.1\,r_{200},r_{200}$).}
\end{center}
\end{minipage}
\end{figure*}

\begin{figure*}
\begin{minipage}{18cm}
\begin{center}
  \includegraphics{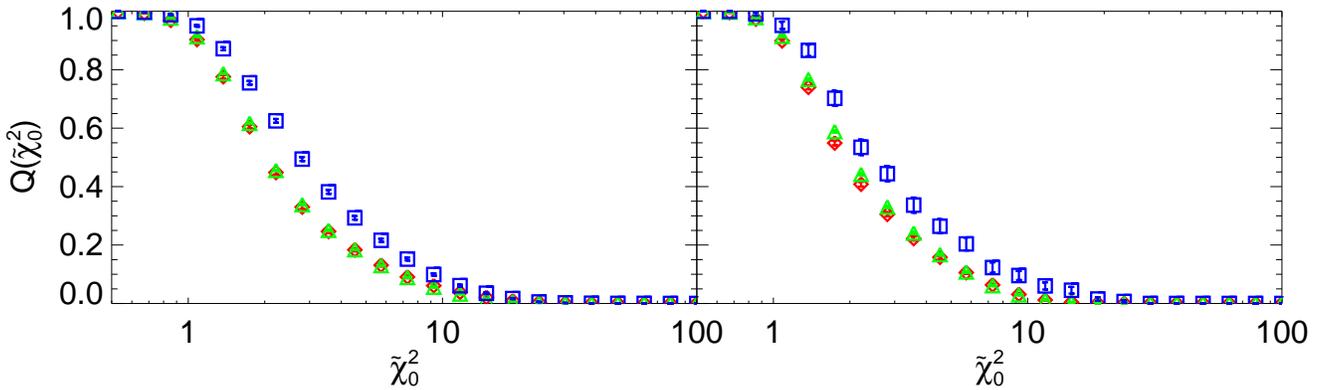}
  \caption{\label{mass-dependancy}$Q(\tilde{\chi}^2)$ for mass bins $10^{12}<M_{200}(\rmn{h^{-1}M_\odot})<10^{13}$ (red diamonds), $10^{13}<M_{200}(\rmn{h^{-1}M_\odot})<10^{14}$ (green triangles) 
and $M_{200}>10^{14}~h^{-1}\rmn{M_\odot}$ (blue squares) in the $\Lambda$CDM-W5 model at $z=0$ (left panel) and $z=1$ (right panel) from the combined halo catalogs of the $512^3$ and $1024^3$ particles simulations.}
\end{center}
\end{minipage}
\end{figure*}

A visual example of this classification is shown in Figure~\ref{fits} where we plot for the the $\Lambda$CDM-W5 model the density profile of nine halos well fit by the 
NFW at more than $2\sigma$ (left panels), at $\sim 1\sigma$ (middle panels) and poorly fit at more than $2\sigma$ (right panels) with masses 
corresponding to galaxy (top panels), groups (central panels) and clusters (bottom panels) halos respectively. Halos with mass $M_{200}>10^{14}$ h$^{-1}\rmn{M_\odot}$ are from
the $1024^3$ simulation, while those with $M_{200}<10^{14}$ h$^{-1}\rmn{M_\odot}$ are from that with $512^3$ particles. The red solid line is the best-fit NFW profile in the interval
($0.1\,r_{200},r_{200}$). We may notice that halos with $\tilde{\chi}^2\lesssim 3$ have profiles that reproduce the NFW formula over the entire radial range. 
In contrast, halos which depart from NFW at more than $2\sigma$ have profiles that are perturbed especially in the external part ($r>0.1\,r_{200}$) 
where the slopes deviates multiple times from that of the NFW profile. This is clearly evident in the case of the most massive halo shown in the right bottom panel of Figure~\ref{fits}. 
The trend inferred from these nine halos is well summarized in Figure~\ref{mass-dependancy}, where we 
plot the cumulative distribution for the halos in the $512^3$ and $1024^3$ particle simulations in the same bins of mass at $z=0$ (left panel) and $1$ (right panel). 
We can see that for both redshifts $Q(\tilde{\chi}^2_0)$ is systematically shifted to larger $\tilde{\chi}^2_0$-values for increasing halo masses.

Large deviations from the NFW profile can be related to the distribution of Dark Matter particles
that have yet to virialize. As recently shown by the work of \citet{Ludlow2012}, this is more likely to occur in the case of massive halos, which have formed 
relatively recent and therefore may still be out-of-equilibrium. 

\begin{figure}
  \includegraphics{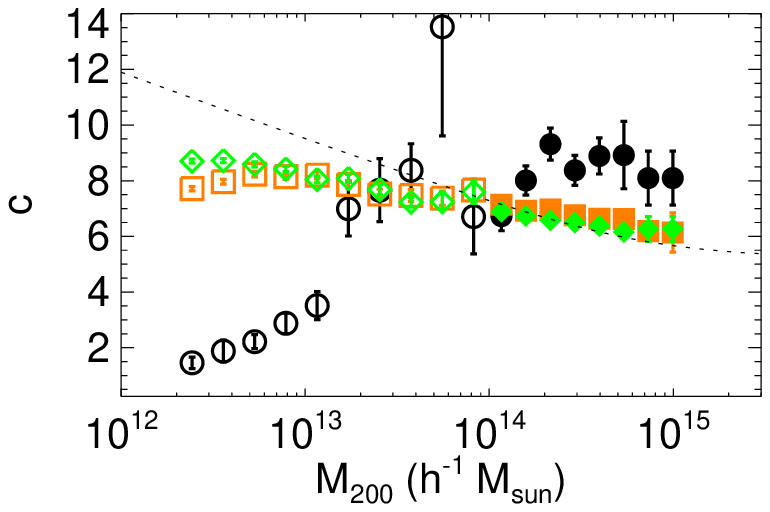}
  \caption{\label{concentration-lcdmw5-0}Mean NFW concentration as a function of halo mass for the $\Lambda$CDM-W5 model simulation at $z=0$. Filled symbols correspond to data points obtained from the $648~h^{-1}$~Mpc boxlength simulation with $1024^3$ particles, while empty symbols correspond to the $162~h^{-1}$~Mpc boxlength simultion with $512^3$ particles. We divide the total subsample in three subsamples: halos with $\tilde{\chi}^2 <3$ (green diamonds, fitted by NFW to better than $1\sigma$), halos with $\tilde{\chi}^2 <10$ (orange squares, fitted by NFW to better than $2\sigma$), halos with $\tilde{\chi}^2 >10$ (black circles, ill-fitted by NFW at more than $2\sigma$).}
\end{figure}

Deviations from the virial condition are 
usually estimated in terms of the parameter $\eta= 2K/|U| - 1$, where $K$ is the kinetic energy and $U$ the potential energy of the halo. However, while
there is no ambiguity in the measurement of the kinetic energy associated with halo particles, the computation of the potential energy is much more subtle.
This is because $U$ is a non-local quantity, since it also depends on particles that do not belong to the halo, but are in the surrounding density field. 
Thus, the evaluation of $\eta$ may be particularly sensitive to the detection algorithm and lead to the definition of a virial selection criterion that 
strongly depends on the halo finder. In Appendix~\ref{etachitilde} we show that indeed the distribution of $\tilde{\chi}^2$ values as function of $\eta$ strongly depend on the halo detection algorithm. Hence, a rigorous assessment of this problem requires a systematic study that at the moment is still missing. It is beyond the scope of this work to investigate the exact dynamical origin of the deviations from NFW and its relation to the goodness-of-fit. It is reasonable to expect that such a correlation exists, but as shown in Appendix~\ref{etachitilde} its significance strongly depends on the halo detection algorithm and the virial selection criteria. However, whether perturbed halo profiles are caused by incomplete-relaxation or dynamical interactions is irrelevant to the point we want to make here. In fact, independently of the dynamical state of halos, it is clear that
if NFW is a poor fit to the halo profile then the information encoded in the concentration parameter is uninformative. To show this we 
plot in Figure~\ref{concentration-lcdmw5-0} the mean halo concentration as function of the total mass for halos which are within $1\sigma$ (green diamonds), $2\sigma$ (orange square) of NFW and ill fitted at more than $2\sigma$ (black circles), errorbars are given by the standard deviation among halos in the same mass bin. We can see that the mean concentration parameter is a monotonically decreasing function of halo mass only for those halos 
whose profiles is within $2\sigma$ of NFW. This is consistent with the picture
that more massive halos form at later times when the mean cosmic density is lower and thus are less concentrated compared to small mass halos which have formed
earlier. In Figure~\ref{concentration-lcdmw5-0} we also plot the concentration predicted by \citet{Zhao2009}, notice that while this model recovers the mean 
concentration of halos well fitted by NFW with masses $M_{200}\gtrsim 6 \times 10^{13}~h^{-1}$~$M_\odot$, at smaller masses it tends to slightly over predict the mean concentration\footnote{Throughout this article, we consistently use the mean as opposed to the median usually considered in the literature, see e.g. \citet{Zhao2009}}.

In the case of halos ill fitted by NFW at more than $2\sigma$ we can see a very different trend, the mean concentration is small at low masses and increases as function of mass
till saturating in the high mass end. Note that the point corresponding to ill-fitted halos at $M_{200} \sim 6 \times 10^{13}$~M$_\odot$ encompasses a very small number of halos. That is why we have a large dispersion and discrepancy in the value of the concentration.

\begin{figure}
\begin{center}
  \includegraphics{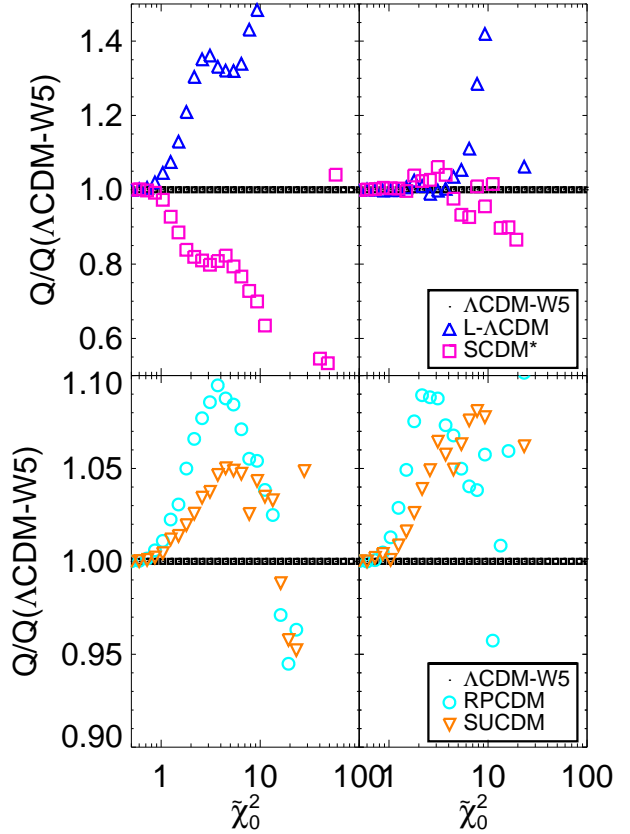}
  \caption{\label{cdfratio_all}Ratio of $Q(\tilde{\chi}^2_0)$ relative to that of the $\Lambda$CDM-W5 cosmology for toy models (top panels) and realistic cosmologies (bottom panels)
at $z=0$ (left panels) and $z=1$ (right panels) respectively.}
\end{center}
\end{figure}

We now focus on the cosmological dependence of the goodness-of-fit. To this purpose in Figure~\ref{cdfratio_all} we plot $Q(\tilde{\chi}^2_0)$ relative to that of the reference $\Lambda$CDM-W5 cosmology
for halos at $z=0$ (left panels) and $z=1$ (right panels) in the case of toy models (top panels) and realistic ones (bottom panels). We can see that for the toy models
at $z=0$ the fraction of halos with $\tilde{\chi}^2>\tilde{\chi}^2_0$ relative to the $\Lambda$CDM-W5 is larger (smaller) for the L-$\Lambda$CDM (SCDM*). 
Differences are smaller at $z=1$ especially in the case of the SCDM*; this is not surprising since at higher redshifts the expansion history of Dark Energy models approaches that of the SCDM* model.
In the case of the realistic models differences in the cumulative distribution function are smaller than for toy models, again this is not surprising since their expansion histories 
are calibrated against the same cosmological dataset, nonetheless we can still notice differences $\lesssim 5\%$ at $z=0$ and of few percent at $z=1$. 
This implies that the population of halos which depart from the NFW profile also carry a distinct signature of the underlying cosmological model. As shown in Figure~\ref{concentration-lcdmw5-0}, whatsoever the dynamical cause of the perturbed density profile, such information is not correctly encoded in the NFW concentration parameter. 

At this point it is reasonable to ask whether assuming a different halo profile fitting formula may lead to different results. Recent studies \citep[see e.g.][]{Ludlow2013} have shown that ``relaxed'' halos in N-body simulations are better described by the Einasto profile \citep{Einasto} rather than NFW. The former is characterized by an additional free parameter. However, the improvement of the fit mainly concerns the core region of halos. In contrast, indepedently of the dynamical state of halos, our analysis highlights deviations from NFW at higher radii, such as those shown in Figure~\ref{fits}, where the Einasto profile closely resembles NFW. Hence, this calls for a characterization of the halo mass distribution that is independent of any parametric fitting formula of the profile.

\section{Halo Sparsity}\label{spars}
\subsection{Definition}
We introduce the halo sparsity defined as the ratio of the halo mass measured at two different overdensities $\Delta_1$ and $\Delta_2$,
\begin{equation}
s_{\Delta_1\Delta_2}\equiv\frac{M_{\Delta_1}}{M_{\Delta_2}},
\end{equation}
with $\Delta_1<\Delta_2$. The sparsity provides an estimate of the halo mass excess contained between the radius $r_{\Delta_2}$ and $r_{\Delta_1}$ relative to the halo
mass enclosed in the inner radius $r_{\Delta_2}$. Hereafter, we fix $\Delta_1=200$ and let $\Delta_2=\Delta$ to vary. However, the general properties of the sparsity which will
be derived here are independent of the specific choice of $\Delta_1$ provided its value is not too small such that the halo retains its individuality. 
Hence, a lower limit on  $\Delta_1$ is probably of order $100$. Conversely, $\Delta_2$ cannot be too large, in which case it will be sensitive to mass distribution inside 
the halo core where baryonic processes, which are not considered here, become relevant. In such a case upper limits on $\Delta_2$ may vary in the range $3000$ to $5000$ depending 
on the total halo mass, redshift and cosmology. Notice that the sparsity is a directly measurable quantity that can be inferred from halo mass observational measurements.

At larger overdensities SO halos are characterized by a smaller number of particles, hence in order to be conservative for $\Delta>200$ we only consider halos with no less than $200$ particles.

\begin{figure}
  \includegraphics{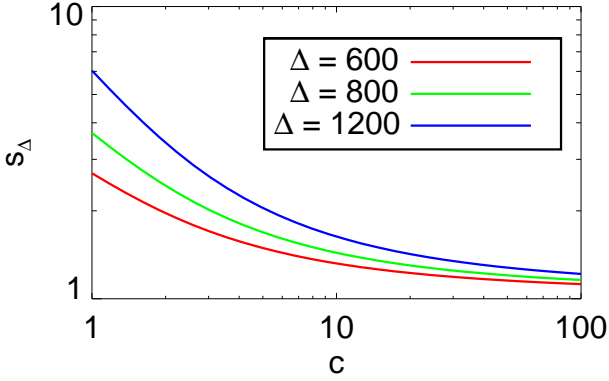}
  \caption{\label{sparsity-concentration} Sparsity as function of the concentration for different overdensity values. The larger the concentration the smaller the sparsity.}
\end{figure}

\subsection{Sparsity vs NFW concentration}
In the case of halos well fit by NFW the sparsity can be related to the concentration parameter. Let $M_\Delta=\frac{4}{3}\pi r_\Delta^3 \Delta \rho_m$ be the mass enclosed in the radius $r_\Delta$. 
We can rewrite the sparsity as $s_\Delta =200/(x^3\Delta)$, where $x=r_\Delta/r_{200}$, then using the NFW formula we find
\begin{equation}
x^3 \frac{\Delta}{200}=\frac{\ln (1+cx) - \frac{cx}{1+cx}}{\ln (1+c) - \frac{c}{1+c}}.\label{spc}
\end{equation}
This equation can be solved numerically to find the relation $s_\Delta(c)$, which we plot in Figure~\ref{sparsity-concentration} for different values of $\Delta$. We 
can see that sparsity and concentration are anti-correlated, which explains our choice of dubbing $s_\Delta$ as sparsity. 

\begin{figure}
\includegraphics{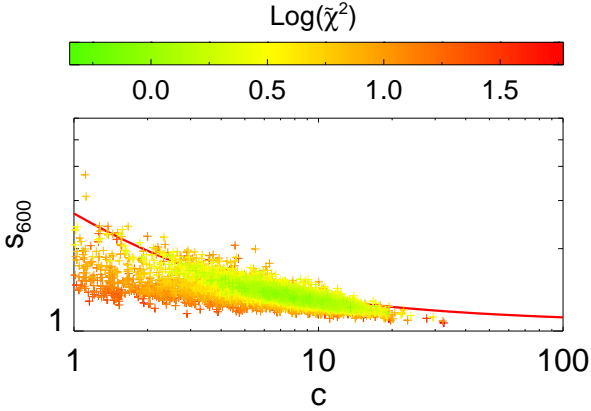}
\caption{\label{sparsity-lcdmw5-z0} Plot of $s_{600}-c$ for halos in the $\Lambda$CDM-W5 cosmology at $z=0$. The red solid line shows the theoretical prediction for a NFW profile. Many halos depart from this prediction as they are not well described by the NFW profile.}
\end{figure}

In Figure~\ref{sparsity-lcdmw5-z0} we plot $s_{600}$ as function of the concentration for halos in the $\Lambda$CDM-W5 catalog. As we can see halos
which are well fit by the NFW profile ($\tilde{\chi}^2_0<3$) have a sparsity that is narrowly distributed along the value predicted by Eq.~(\ref{spc}). This is clearly
not the case for halos whose profile departs from NFW, which is an indication that the concentration parameter (i.e. NFW profile) does not correctly track the mass distribution 
in the external part of halos between $r_{600}$ and $r_{200}$. Here, it is worth noticing that over the entire mass range of the halo catalog
the sparsity is characterized by a much smaller dispersion than the concentration parameter.
   
\begin{figure}
\begin{center}
  \includegraphics{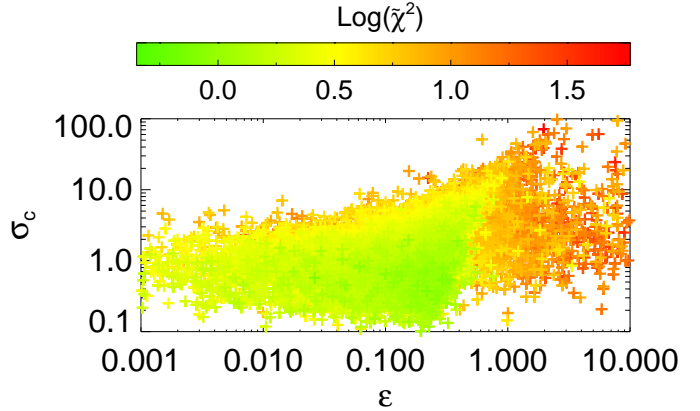}
\caption{\label{eps-sigma}Dispersion about the average value of the sparsity inferred concentration $\bar{c}$ for each halo in the $\Lambda$CDM-W5 catalog at $z=0$ as function of the deviation from 
the NFW concentration best-fit value, $\epsilon = |\bar{c}- c_\rmn{NFW}|/c_\rmn{NFW}$.}
\end{center}
\end{figure}

We can reconstruct the halo mass profile by measuring the sparsity at several overdensities and compare it to the NFW prediction. 
For each halo in our catalog we measure $s_\Delta$ for $\Delta=\{500,600,800,1000,1200,1500,2000,3000,4000,5000\}$. Then, 
we use Eq.~(\ref{spc}) to derive the corresponding values of $c(s_\Delta)$ from which we compute the average concentration $\bar{c}$ and dispersion $\sigma_c$ of each halo:
\begin{equation}
\bar{c}=\frac{1}{N_\Delta}\sum_{j=1}^{N_\Delta} c(s_{\Delta_j}) \,\,\,\,\,\,\,\,\,\,\,\sigma_c=\frac{1}{N_\Delta-1}\sum_{j=1}^{N_\Delta} [c(s_{\Delta_j})-\bar{c}]^2.\nonumber
\end{equation}
In Figure~\ref{eps-sigma} we plot $\sigma_c$ as function of $\epsilon = |\bar{c}- c_\rmn{NFW}|/c_\rmn{NFW}$, where $\epsilon$ measures the difference between the average halo concentration of each halo
inferred from the sparsity relative to the best-fit value of the NFW concentration parameter. We can see a strong correlation as function of $\tilde{\chi}^2$ and $\sigma_c$. In particular, for
halos which are well described by NFW, the sparsity inferred concentration $\bar{c}$ coincides with the best-fit NFW concentration parameter to better than a few percent.
In contrast, halos with $\tilde{\chi}^2\gtrsim 10$ are associated with deviation $\epsilon\gtrsim 0.5$ and scatter $\sigma_c\gtrsim 1$. This means that for such halos the
NFW concentration is no longer representative of the average compactness of the halo, while the correlation with the large values of $\sigma_c$ suggests that such 
deviations are caused by large fluctuations of the mass distribution in halo radial bins.
 
\subsection{Sparsity and Halo Mass}
\begin{figure}
\begin{center}
  \includegraphics{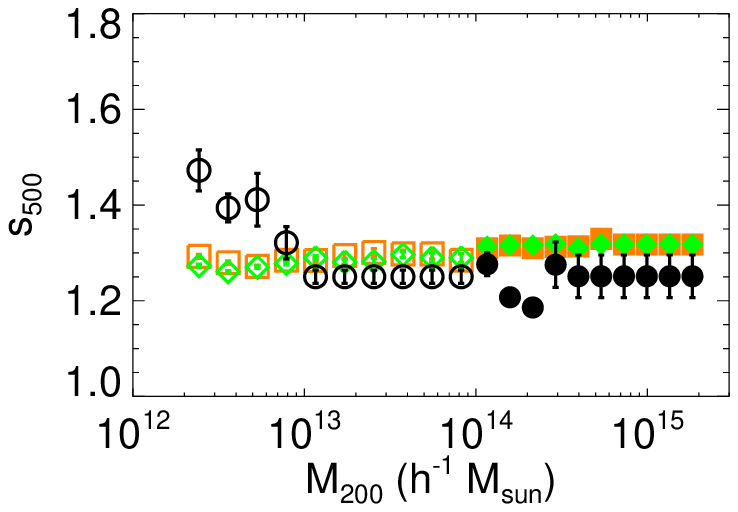}
  \includegraphics{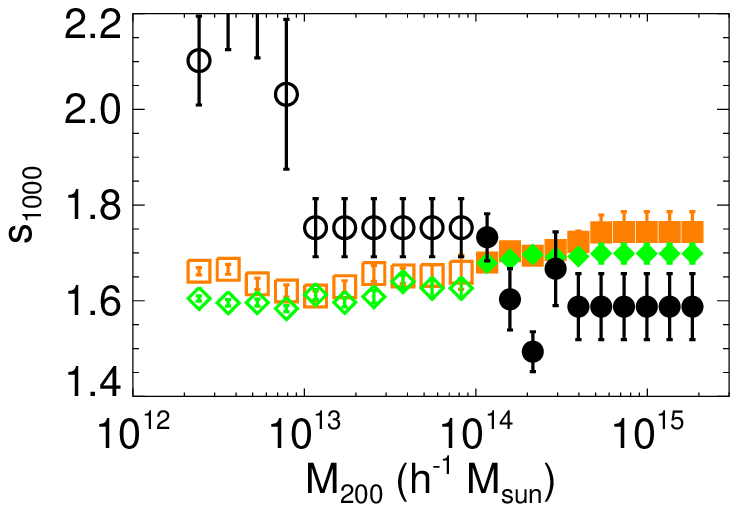}
  \caption{\label{sparsity-M-lcdmw5-z0} $s_{500}$ (top panel) and $s_{1000}$ (bottom panel) as function of the total halo mass in the $\Lambda$CDM-W5 cosmology at $z=0$. Filled symbols correspond to data points obtained from the $648~h^{-1}$~Mpc boxlength, $1024^3$ particles simulation, while empty symbols correspond to the $162~h^{-1}$~Mpc boxlength, $512^3$ particles simulation. We divide the total subsample in three subsamples: halos with $\tilde{\chi}^2 <3$ (green diamonds, fitted by NFW to better than $1\sigma$), halos with $\tilde{\chi}^2 <10$ (orange squares, fitted by NFW to better than $2\sigma$), halos with $\tilde{\chi}^2 >10$ (black circles, ill-fitted by NFW). The sparsity for all categories of halos now follows the same trend.}
\end{center}
\end{figure}

We find the sparsity to be nearly independent of the total halo mass. In Figure~\ref{sparsity-M-lcdmw5-z0} we plot the average value of the sparsity as function of halo mass for
$\Delta=500$ (top panel) and $\Delta=1000$ (bottom panel) respectively with one standard deviation errorbars for halos which are within $1$ (green diamonds) and $2\sigma$ (orange squares)
of the NFW profiles and ill fitted at more than $2\sigma$ (black circles). For $M_{200}<10^{14}$ h$^{-1}\rmn{M_\odot}$ the sparsity is computed using halos 
in the $512^3$ simulation, while for larger masses we use the $1024^3$ halo catalog. We can see that $\langle s_{500}\rangle$ varies less than $5\%$ over the entire
mass range, independently of whether the halos are well described by the NFW profile. At low masses the scatter is mainly due to resolution issues since for these halos the number of particles
is close to the minimum value of $200$. For $\Delta=1000$ the dependence on the total halo mass is slightly more accentuated especially in the high-mass
end with variations $\lesssim 10\%$. In the case of halos ill fitted by NFW the scatter in the value of the sparsity is also within the $10\%$ level, contrary to what was observed for the concentration in figure~\ref{concentration-lcdmw5-0}. The high value and dispersion observed for low mass halos is a resolution effect: the mass $M_\Delta$ is likely underestimated for these halos, as we investigate small radii. This interpretation is consistent with the fact that the discrepancy is larger at higher $\Delta$.

\subsection{Halo Sparsity and Mass Function Consistency Relation}
\label{sparsity-massfunction}
Here we show that the average value of the sparsity can be derived from the halo mass function. Let us write the identity
\begin{equation}\label{smf}
\frac{\rmn{d}n}{\rmn{d}M_\Delta}=\frac{\rmn{d}n}{\rmn{d}M_{200}} \frac{\rmn{d}\ln M_{200}}{\rmn{d}\ln M_\Delta} \frac{M_{200}}{M_\Delta}=s_\Delta \frac{\rmn{d}n}{\rmn{d}M_{200}} \frac{\rmn{d}\ln M_{200}}{\rmn{d}\ln M_\Delta},
\end{equation}
where $\rmn{d}n/dM_\Delta$ and $\rmn{d}n/dM_{200}$ are the mass functions at $\Delta$ and $\Delta=200$ respectively. Since we have shown that $s_\Delta$ is nearly independent of mass, 
the average over the halo ensemble reads as
\begin{equation}
\label{smfconsistency}
 \int^{M_1}_{M_2}\frac{\rmn{d} n}{\rmn{d} \ln M_\Delta} \, \frac{\rmn{d} \ln M_\Delta}{M_\Delta} = \langle s_\Delta \rangle \int^{\langle s_\Delta \rangle M_1}_{\langle s_\Delta \rangle M_2} \frac{\rmn{d} n}{\rmn{d} \ln M_{200}} \, \frac{\rmn{d} \ln M_{200}}{M_{200}},
\end{equation}
the above relation is transcendental in $\langle s_\Delta \rangle$ and can be solved numerically given the mass function at $\Delta=200$ and $\Delta$. 
This implies that knowledge of the mass function at two different overdensities can be used to predict the average value of
the sparsity. Vice versa, since Eq.~(\ref{smfconsistency}) involves all potentially observable quantities, 
it can be used as a consistency test. In fact, let us imagine of observing a sample of clusters for which we measure the mass at two different overdensities,
then the inferred mass functions and the ensemble average of the sparsity must satisfy Eq.~(\ref{smfconsistency})\footnote{A similar consistency test can be inferred by noticing that Eq.~(\ref{smf}) can also be written as  $\langle s_\Delta\rangle=\langle 1/M_\Delta\rangle/\langle 1/M_{200}\rangle$. Since the sparsity is nearly independent $M_{200}$ 
then $\langle s_\Delta\rangle=\langle M_{200}\rangle/\langle M_\Delta\rangle$ must be verified. We tested both relations and found them to be in good agreementt.}.

\begin{figure}
  \includegraphics{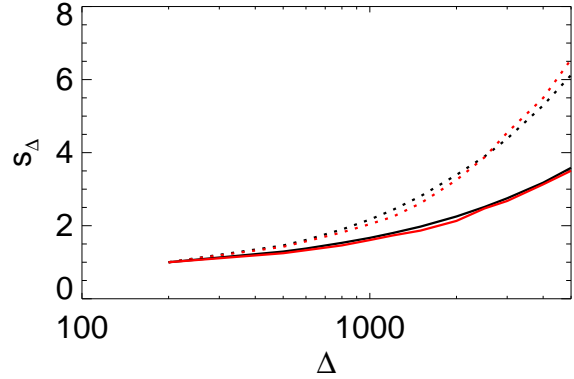}
  \caption{\label{sparsity-measured} $\langle s_\Delta\rangle$ as a function of $\Delta$ at $z=0$ (solid lines) and $z=1$ (dotted lines). 
Black lines correspond to the measured sparsity in the halo catalog, while in red are the mass function based predictions using Eq.~(\ref{smfconsistency}). Both are in remarquable agreement.}
\end{figure}

In Figure~\ref{sparsity-measured} we plot $\langle s_\Delta\rangle$ as a function of $\Delta$ at $z=0$ (solid lines) and $z=1$ (dotted lines) 
computed from the halos in the $\Lambda$CDM-W5 catalog along with the prediction from Eq.~(\ref{smfconsistency}) using the corresponding 
halo mass functions at $\Delta$ and $\Delta=200$. We can see that the mass function based predictions work remarkably well at different 
redshifts and up to large overdensities with residuals $\lesssim 5\%$. 

\section{Cosmology with Halo Sparsity}\label{cosmospar}
We now focus on the imprint of the underlying cosmological model (and specifically of DE) on the halo sparsity. 

\begin{figure}
\begin{center}
  \includegraphics{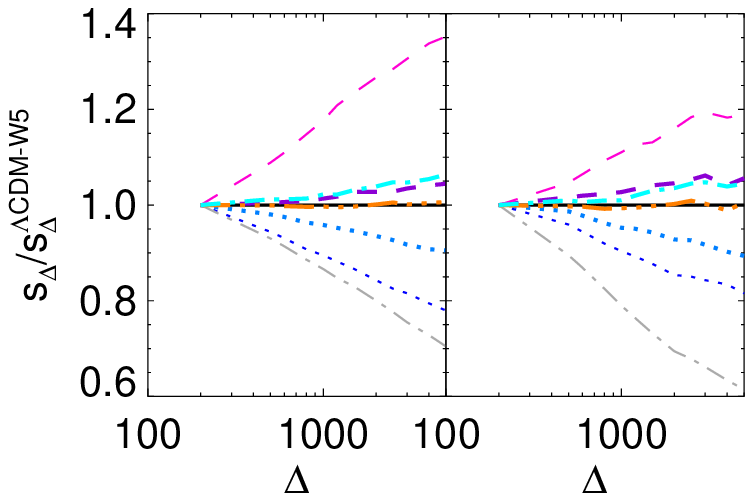}
  \caption{\label{sparsity-cosmo-diff}Sparsity of the simulated cosmological models relative to $\Lambda$CDM-W5 at $z=0$ (left panel) and $z=1$ (right panel) respectively. The bottom panels show a zoom around the $\Lambda$CDM-W5 line. The differences, up to 5\% even in realistic cosmologies, arise from different structure formation histories.}
  \includegraphics{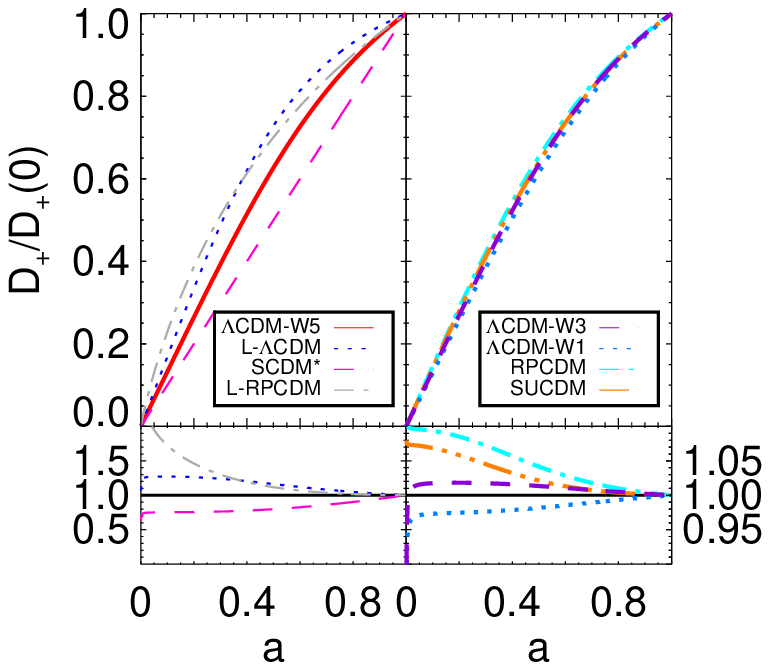}
  \caption{\label{growthrate}Linear growth rate cosmologies for the toy models (left panels) and realistic cosmologies (right panel). 
The bottom panels show the relative difference with respect to $\Lambda$CDM-W5.}
\end{center}
\end{figure}

In Figure~\ref{sparsity-cosmo-diff} we plot the ratio of $\langle s_\Delta\rangle$ relative to that of the reference $\Lambda$CDM-W5 model 
at $z=0$ (left panels) and $z=1$ (right panels) for all models listed in Table~\ref{wmap_cosmo}. The bottom panels show
a zoom around the $\Lambda$CDM-W5 line. We can clearly see that the average sparsity at a given $\Delta$ varies from one model to another. We find such differences to be correlated
with the linear growth histories of the simulated models as well as the value of $\sigma_8$. For comparison in Figure~\ref{growthrate} we plot the linear growth function normalized 
to its present value as function of the scale factor for the simulated cosmologies. In the bottom panels is shown the ratio with respect to $\Lambda$CDM-W5.

Let us first consider the toy models at $z=0$. We can see that in the case of the SCDM* model the average sparsity is up to $\sim 30\%$ larger than in $\Lambda$CDM-W5, 
while in the case of L-RPCDM and L-$\Lambda$CDM this is up to $\sim 30\%$ and $\sim 20\%$ smaller respectively. By construction the toy models have 
the same $\sigma_8$ value. As it can be noticed from the plot in the left panel of Figure~\ref{growthrate} the SCDM* 
linear growth history is suppressed compared to the $\Lambda$CDM-W5, while it is enhanced for L-RPCDM and L-$\Lambda$CDM with the former having the largest enhancement. 
Thus, in the SCDM* case structures form later compared to $\Lambda$CDM-W5, hence on average the mass assembled at large overdensities will be smaller, thus resulting in an enhanced average
sparsity relative to the $\Lambda$CDM-W5. The opposite occurs for L-RPCDM and L-$\Lambda$CDM. 

Let us now consider $\Lambda$CDM-W1 and $\Lambda$CDM-W3 cosmologies. These have nearly identical
linear growth histories (see Figure~\ref{growthrate}, right panel), while they have different values of $\sigma_8$. In particular, $\Lambda$CDM-W1 has the largest value $\sigma_8=0.9$. Because of this, 
it forms structures earlier than $\Lambda$CDM-W5. Consequently, the mass at larger overdensities is greater, which results in a smaller average sparsity compared to that of $\Lambda$CDM-W5. 
In contrast, $\Lambda$CDM-W3 has a slightly smaller value of $\sigma_8$, consequently the average sparsity is larger than $\Lambda$CDM-W5. In the case of RPCDM and SUCDM, the combined effects of the 
linear growth history and $\sigma_8$ compete to give the differences shown in Figure~\ref{sparsity-cosmo-diff}.

These cosmological dependencies are consistent with those expected from Eq.~(\ref{smfconsistency}). In fact, we can factorize the mass function
in terms of the mean cosmic matter density, the derivative of the variance of the linear density field with respect to the mass and the multiplicity function. 
Since the average sparsity is given by the ratio of the integral of the mass function at two different overdensities,
the explicit dependence on the cosmic mean matter density cancels out and remains that on $\sigma_8$ and the linear growth factor. 

A test of the consistency relation presented in Eq.~(\ref{smfconsistency}) yields similar results for all cosmological models, with the error being of order a few percent in each case.

These results are consistent with previous findings \citep[see e.g.][]{Dolag2004,Maio2006,Ma2007,Alimi2010,Courtin2011,DeBoni2013}, namely that the non-linear structure formation still carries a cosmological imprint of the past linear growth history.

Overall, this suggests that the sparsity is a sensitive probe of cosmology. The larger the difference
in the integrated linear growth of density fluctuations and/or their normalization amplitude and the larger the imprint on the sparsity.

\section{Sparsity and Halo Mass Measurements}
Galaxy clusters are host in massive DM halos. These can be detected with a variety
of techniques such as the X-ray emission of the hot intra-cluster gas, the Sunyaev-Zeldovich effects or optical identifications.
In recent years numerous survey programs have provided large sample of galaxy clusters up to high redshift \citep{PLANCK,SPT,MAXBCG,ACT,XXL}. 
A key aspect of these studies is the measurement of the cluster mass. Under certain hypothesis this can be inferred from locally calibrated scaling relations. 
Alternatively the cluster mass can be inferred from gravitational lensing observations
or measurements of the cluster richness. Each of these methods has its own limitations and is affected by different systematic effects.
As an example the hot intra-cluster gas may not be in hydrostatic equilibrium, especially if the cluster is not virialized, this may lead to biased mass estimates.
Gravitational lensing measurements may be more reliable as they probe the entire gravitating mass, on the other hand they are affected by mass perturbations present along the line-of-sight. 

A measurement of the sparsity requires the halo mass at different overdensities. The range of overdensities that can be probed in a cluster depends on the observational method.
As an example X-ray observations cannot probe $\Delta=200$, while they are more likely to measure the mass at $\Delta=500$. Higher overdensities require sufficient angular resolution. 
The inferred value of the sparsity can then suffer of systematic uncertainties affecting the mass measurements. Nevertheless, the existence of the sparsity consistency relation shown in Eq.~(\ref{smfconsistency})
on a large cluster sample can provide an effective diagnostic for testing unknown systematics. 

\begin{figure}
  \includegraphics{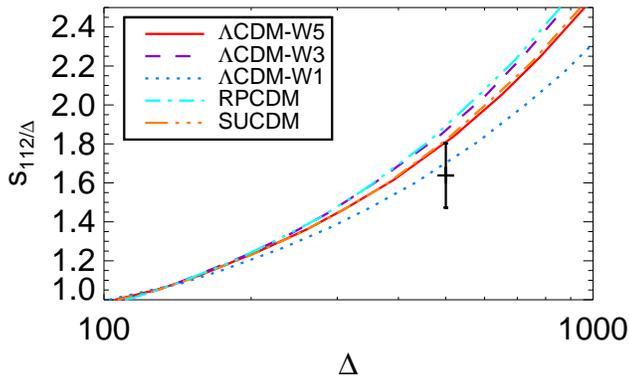}
  \caption{\label{s_obs} Comparison of the average sparsity of a cluster sample with 2D lensing masses from \citet{Okabe2010} against predictions from the simulated cosmological models. The overdensity $\Delta$ quoted here is with respect to the critical density.}
\end{figure}

To give an illustrative example of the cosmological relevance of the sparsity we compute the average value from a sample of 30 clusters around $z\sim 0.2$ from \citet{Okabe2010} (see table~1 of this article for the redshifts and table~8 for the 2D masses) for which 
lensing mass estimates at overdensities $\Delta_c=112$ and $\Delta_c=500$ with respect to the critical density have been obtained without assuming a NFW profile. These 
are 2D projected masses and in principle we should compare it to the sparsity of 2D halo masses from our catalogs. However, if we assume the halos to be approximately spherical, and given the fact that we study the ratio of two masses, we expect the differences to be minimal. The average sparsity of the cluster sample is 
$\langle s_{112,500}\rangle=1.71\pm 0.38$ which we plot in Figure~\ref{s_obs} together with the simulated cosmological model predictions. The latter have been obtained by converting the N-body halo mass 
measurements with respect to the critical density using the relation $\Delta_c = \Omega_m\Delta (1+z)^3/E^2(z)$, where $E^2(z)=(1-\Omega_m)+\Omega_m(1+z)^3$ in the case of a flat universe (having neglected the contribution of radiation). In this proof-of-concept, we use the simulations at $z=0$ to compare with the observations at $z\sim 0.2$ since we lack the simulation snapshots at the correct redshift. 

As we can see even a single estimate of the sparsity can potentially have a significant constraining power and which is worth
to further investigate in future work.

\section{Conclusions and perspectives}\label{conclu}
In this paper we have studied the imprint of DE on the density profile of DM halos using a set of halo catalogs from the DEUSS project.
We have shown that the goodness-of-fit of the NFW profiles varies with mass, redshift and cosmology. 
In particular, the fraction of halos ill fitted by NFW systematically
varies with the underlying cosmologies. Thus, the mass distribution in such halos still carries cosmological information. 

We have introduced the halo sparsity to retrieve the DE dependent signature encoded in the mass distribution inside all halos independently of a parametric profile. 
We have shown that the average value of the sparsity is related to the halo mass function at different overdensities and we have inferred a consistency
relation that can be used either to predict its value if the halo mass functions are known or validate observational measurements of the sparsity.

In the future large sample of galaxy clusters from multi-wavelength observations can be used to infer accurate measurements of the sparsity from which
it will be possible to derive alternative cosmological constraints.

\section*{Acknowledgements}
I. Balm\`es is supported by a scolarship of the ``Minist\`ere de l'\'Education Nationale, de la Recherche et de la Technologie'' (MENRT). The research leading to these results has received funding from the European Research Council under the European Community's Seventh Framework Programme (FP7/2007-2013 Grant Agreement no. 279954). This work was granted access to the HPC resources of CCRT and IDRIS under allocations made by GENCI (Grand \'Equipement National de Calcul Intensif).

\appendix

\section{FoF Halos}\label{fofhalos}
In this Appendix we briefly present the results on the halo profiles obtained from halos detected 
with the FoF algorithm with $b=0.2$. The purpose is to show that the results on the NFW goodness-of-fit and
the halo sparsity do not depend on the precise identification criterion of halos.
Here we restrict ourselves to the results obtained with the $162~h^{-1}$~Mpc boxlength and
$512^3$ particles simulation. Note that the mass indicated for halos is $M_{200}$, i.e. the mass
obtained by constructing a sphere of overdensity $200$ around the density peak of the
structure found by the FoF algorithm. 
We fit the density profile of FoF halos using the same procedure used for
SO ones, and use $\tilde{\chi}^2$ to evaluate the goodness-of-fit. 

\begin{figure}
\begin{center}
\includegraphics{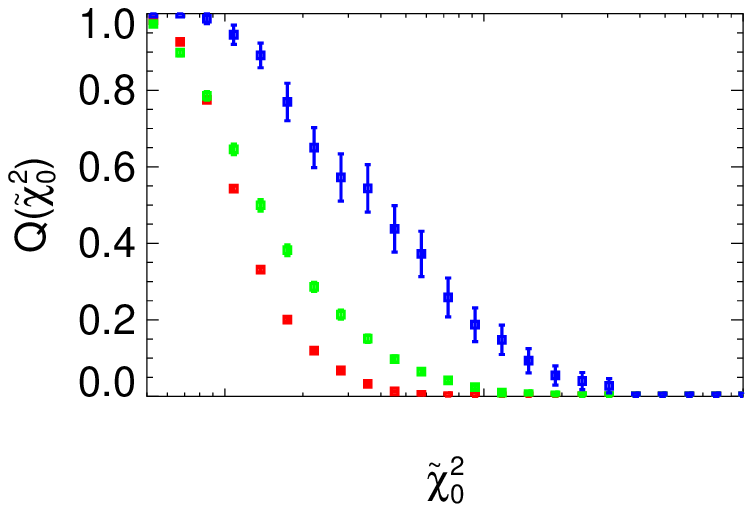}
\caption{\label{appendix-chisqmass}$Q(\tilde{\chi}_0^2)$ for FoF halos at $z=0$ in different bins of mass.}
\end{center}
\end{figure}

Figure~\ref{appendix-chisqmass} shows the cumulative distribution function, $Q(\tilde{\chi}_0^2)$, obtained from
FoF halos respectively for three different bins of mass. This figure is to be compared with the left panel of Figure~\ref{mass-dependancy}.
The dependence on the halo mass is more important for FoF halos, possibly due to a 
remaining dependence of the goodness-of-fit on the mass resolution. 
Surprisingly, FoF halos seem to have a generally lower $\tilde{\chi}^2$ value 
than SO halos, especially at small masses. This is likely due to the rescaling factor
we apply to ${\chi}^2$ to account for resolution effects. A more careful study is needed
to determinate the appropriate scaling in the case of FoF halos. 

In Table~\ref{appendix-chisqsigma} we list the values of $\tilde{\chi}^2$ corresponding to $1$ and $2\sigma$ deviations from
the NFW profile. For SO halos, we have obtained values around $3$ and $10$ respectively,
while in the case of FoF halos these are around $1$ and $3$. 

\begin{table}
\begin{center}
\caption{\label{appendix-chisqsigma}Values of $\tilde{\chi}^2$ corresponding to 1 and 2$\sigma$ deviation from the NFW profile for FoF halos in the simulated cosmologies at $z=0$ and $1$ respectively.}
\begin{tabular}{ccccc}
\hline
& \multicolumn{2}{c}{$z=0$} & \multicolumn{2}{c}{$z=1$} \\
& 1$\sigma$ & 2$\sigma$ & 1$\sigma$ & 2$\sigma$ \\ \hline
$\Lambda$CDM-W5 & $1.49$ & $3.86$ & $1.55$ & $3.16$ \\
RPCDM  & $1.56$ & $4.05$ & $1.62$ & $3.31$ \\
SUCDM & $1.51$ & $4.06$ & $1.56$ & $3.19$ \\
L-$\Lambda$CDM & $1.70$ & $4.68$ & $1.49$ & $3.29$ \\
SCDM*  & $1.42$ & $3.16$ & $1.63$ & $3.26$ \\
\hline
\end{tabular}
\end{center}
\end{table} 

\begin{figure}
\begin{center}
\includegraphics{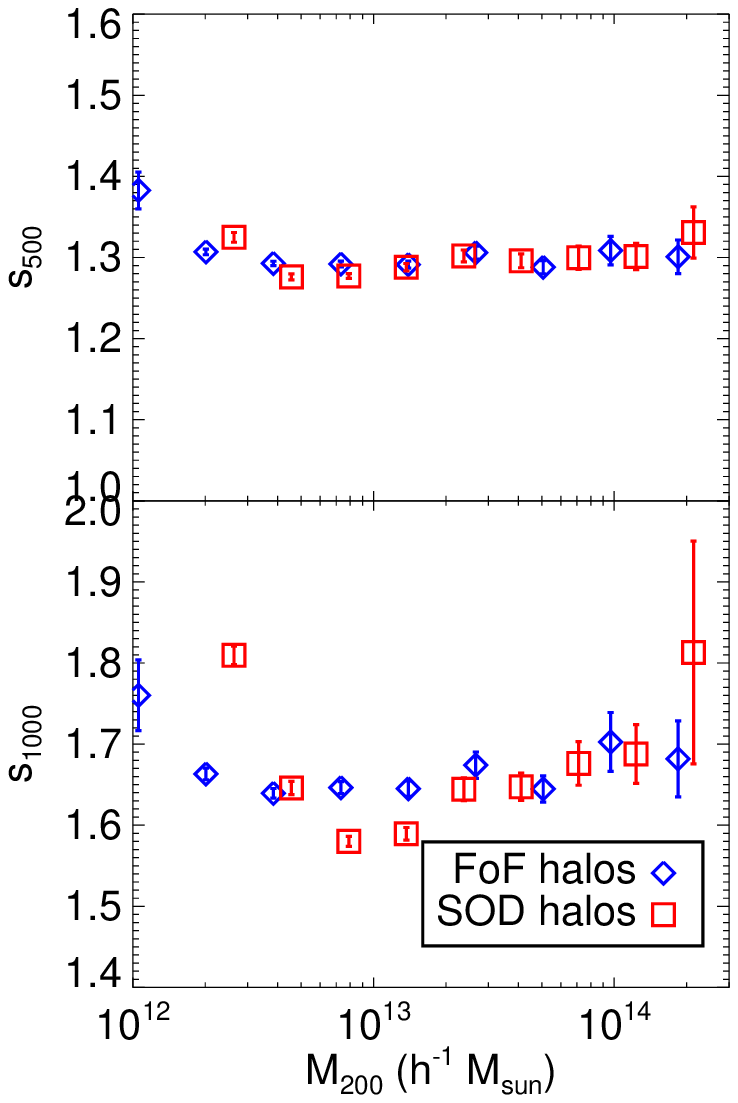}
\caption{\label{appendix-sparsity}Comparison of $s_{500}$ and $s_{1000}$ derived from FoF and SOD halos in $\Lambda$CDM-W5.}
\end{center}
\end{figure}

In Figure~\ref{appendix-sparsity} we plot the halo sparsity at $\Delta=500$ and $1000$ respectively for both SO and FoF halos, 
we can see that the results are remarkably similar. 

\section{Goodness-of-fit vs Dynamical State}\label{etachitilde}
Here, we compute a diagnostic of the dynamical state of halos as measured by the parameter $\eta= 2K/|U| - 1$, where $K$ is the kinetic energy and $U$ is an estimate of the potential energy of the halo. Values of $\eta\sim\mathcal{O}(1)$ are indicative of deviation from the virial theorem. For instance \citet{Neto2007} consider halos to be virialized if $\eta<0.3$, while \citet{Ludlow2013} assume a more stringent criterion $\eta<0.1$. Other proxies used in combination with $\eta$ are the measure of the off-set between the peak density of the halo and its center of mass. 

As pointed out in Section~\ref{chi2}, while $K$ is unambigously defined in terms of the halo particles (since it is given by the sum of the kinetic energy of all particles in the halo). In contrast $U$ is a non-local quantity which is usually estimated by computing the potential energy between pairs of particles in the halo. However, this only gives a lower limit to the total potential energy since there could be non-negligible contributions from particles residing in the surrounding density field or nearby halos. This implies that results may strongly depend on the halo detection algorithm or the virial selection criteria. 

\begin{figure}
\begin{center}
  \includegraphics[width=0.95\hsize]{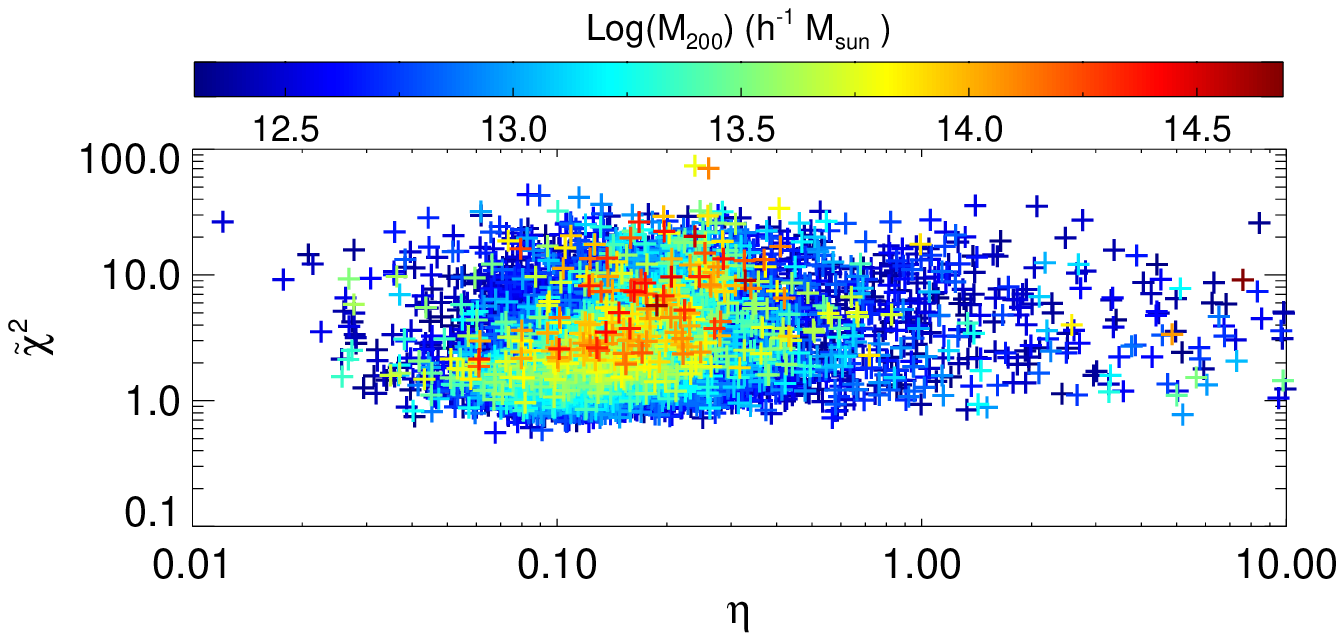}
  \includegraphics[width=0.95\hsize]{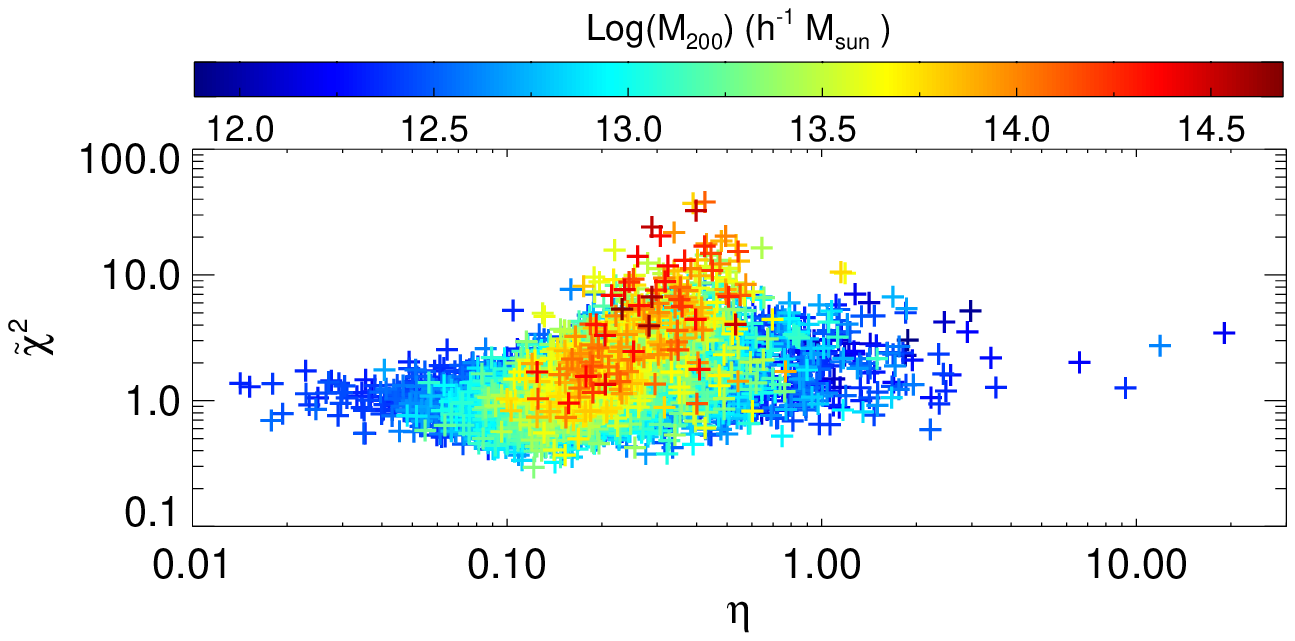}
  \caption{\label{chi2vir_so_fof}$\tilde{\chi}^2$ as function of $\eta$ for SO (upper panel) and FoF (lower panel) halos at $z=0$ from the $\Lambda$CDM-W5 simulation with $512^3$ particles and $162$~h$^{-1}$ Mpc boxlength.}
\end{center}
\end{figure}

In the upper panel of Figure~\ref{chi2vir_so_fof} we plot the distribution of $\tilde{\chi}^2$-values as function of $\eta$ at $z=0$ for the SO halos from the $\Lambda$CDM-W5 simulation with $512^3$ particles and $162$~h$^{-1}$ Mpc boxlength. We can see that for the most massive halos ($M_{200}\gtrsim 2\times 10^{14}$~h$^{-1}$~M$_\odot$) $\tilde{\chi}^2$ is an increasing function of $\eta$. If we assume halos to be virialized for $\eta<0.3$ then there is a non-negligible fraction of halos which are more than $2\sigma$ from NFW ($\tilde{\chi}^2\gtrsim 10$) and deviates from the virial condition. Instead if we consider the more stringent criterion $\eta<0.1$ the fraction reduces accordingly. In any case the correlation remains difficult to asses because of a large scatter. This is not the case of FoF halos from the same simulation which are shown in the lower panel of Figure~\ref{chi2vir_so_fof}. Here, a correlation between $\tilde{\chi}^2$ and $\eta$ clearly stands out for the most massive halos. We can see that a large fraction of them are more than $2\sigma$ from NFW ($\tilde{\chi}^2\gtrsim 4$) while deviating from ``equilibrium'' even assuming a less conservative criterion $\eta<0.3$. 

The difference between SO and FoF analyses shows that it is far from trivial to establish the dynamical state of halos in absolute terms since the diagnostic critically depends on the halo detection algorithm and the selection criteria\footnote{We find similar results for $\tilde{\chi}^2$ as function of the halo off-set which we do not show here.}.

This analysis indicates that the deviations from NFW are correlated with the dynamical state halos, but the significance of this remains difficult to asses, thus requiring a dedicated separate study that goes beyond the scope of the work presented here.

\end{document}